\documentclass[article]{aa}

\usepackage{graphicx}
\usepackage{ae}
\usepackage{txfonts}
\usepackage{natbib}
\bibpunct{(}{)}{;}{a}{}{,}

\begin{document}

\title {The Galactic thick and thin disks: differences in evolution }  
 \author{Tetyana\ V.\ Nykytyuk\inst{1} 
 \and Tamara\ V.\ Mishenina\inst{2}} 

\institute{Main Astronomical Observatory, Ak.Zabolotnoho St.27, Kyiv 03680, 
Ukraine 
\and Astronomical Observatory Odessa National University, 
Odessa, Ukraine}

\authorrunning {T.V.\  Nykytyuk et al.} 
\titlerunning {The Galactic thick and thin disks: differences in evolution} 
\offprints {T.V.\ Nykytyuk, \email {nikita@mao.kiev.ua}}

%\em nikita@mao.kiev.ua}

%{\em tamar@deneb.odessa.ua}

\date{Received <date> / Accepted <date>}

\abstract{Recent observations demonstrate that the thin and thick disks of 
the Galaxy have different chemical abundance trends and evolution timescales.} 
{The relative abundances of $\alpha$-elements 
in the thick Galactic disk are increased relative to the 
thin disk. Our goal is to investigate the cause of such differences in 
thick and thin disk abundances.} 
{We investigate the 
chemical evolution of the 
Galactic disk in the framework of the open two-zone model with gas inflow.}
{The Galactic abundance trends for  
$\alpha$-elements (Mg, Si, O) and Fe 
are predicted for the thin and thick Galactic disks.}
{The star formation histories 
of the thin and thick disks must have been different and the gas infall 
must have been more intense during the thick disk evolution that 
the thin disk evolution.}      


\maketitle

\section{Introduction}

Recent studies of the kinematics and ages of disk stars of our Galaxy 
have revealed the presence of two distinct  
populations in the Galactic disk, called the thick and the thin disk 
populations. 
\citet{GR83} have shown that the stellar density distribution 
towards the Galactic South Pole requires the presence of at least two kinematically 
distinct disk structures. 
For z<~1 kpc their data are fitted by a single exponent 
and  by a second exponent in the range 1 - 5 kpc above Galactic plane.  
The authors considered this as evidence of the presence of an old thin disk and 
a thick disk. 
The thick disk stars have a scale height from 760 to 1450 pc
\citep{GR83,R96,RR01,C97}; 
the scale height of the thin disk is 100 to 340 pc \citep{GR83,C97}.
The velocity dispersion ($\sigma_u,\sigma_v,\sigma_w$) of the thick and 
thin disk stars is also different; it is 
(63$\pm$6,39$\pm$4,39$\pm$4) km/s and 
(39$\pm$2,20$\pm$2,20$\pm$1) correspondingly \citep{Sou03}.

\citet{J98} used the Hipparcos color-magnitude diagram of 
field stars and derived a minimum age for the  Galactic disk of about 8 Gyr.  
\citet{Fuh98} found that the thin disk stars in his sample 
are younger than 8-9 Gyrs while
the thick disk star ages exceed 10 Gyrs.  
\citet{BFF01} suggested that the maximum age  for the thin disk stars is near 
9 Gyrs whereas the thick disk stars have ages from 12 to 14 Gyrs. 
\citet{BFL03} found that 
the thick disk stars are older on average than the thin disk stars,
$11.2\pm4.3$ Gyrs and $4.9\pm2.8$ Gyrs.
In \citet{FBL03} they derived slightly higher average ages -
$12.1\pm3.8$ Gyrs and $6.1\pm2.0$ Gyrs for thick and thin disks correspondingly.
Thus, the thin and thick disks were formed at different 
epochs.

The chemical characteristics of these two disk subsystems of the Galaxy
are also noticeably different.
A kinematic study of the local G dwarf metallicity distribution of our Galaxy 
provides evidence that the abundance distribution below 
[Fe/H]$\sim$ -0.4 contains two overlapping distributions, the thin and the 
thick disks \citep{WG95}.
The mean metallicity of the thick disk population, <[Fe/H]>, covers  
the range between  -0.5 and -0.7 dex \citep{Sou03,WG95,R96}
while the mean value of the thin disk star metallicity is -0.17 to -0.25 dex 
\citep{Sou03,WG95}.

The thick and thin disk abundance trends partly 
overlap in the range -0.8 < [Fe/H] < -0.4 
but they are different in [$\alpha$/Fe].
\citet{Gr96} pointed out that changes in
[Fe/O] values of the disk stars can be explained by a systematic 
difference in the chemical composition of the thin and thick disk stars.
Later \citet{Gr00} studied this interpretation in detail; as their [Fe/O]vs[O/H] diagram
shows, there are two groups of disk stars with [O/H] > -0.5 -
the thin disk stars with  [Fe/O] > -0.25 and the thick disk stars 
with [Fe/O] < -0.25.

The thick disk has a higher [O/Fe] ratio than the thin disk at sub-solar 
metallicities. The thick disk also shows signatures of chemical 
enrichment  by type Ia supernovae \citep{BFL04}.
\citet{Fuh98} have demonstrated 
a clear disk separation in [Mg/Fe] in the range -0.6 < [Fe/H] < -0.3
in their [Mg/Fe]vs[Fe/H] diagram - the stars with thick disk kinematics have 
[Mg/Fe]$\simeq$+0.4 while the thin disk stars have [Mg/Fe] value decreasing 
from +0.2 to 0.0 dex. \citet{BFL03} have analyzed the  
spectra of F and G  disk dwarfs with metallicities located 
in the range -0.8 < [Fe/H] < +0.4 and have found that the thin and thick 
disk abundance trends are clearly separated at [Fe/H] < 0.
\citet{BFL05} confirms their previous results - 
there are  distinct and separate abundance trends between the thin and 
thick disks. The thick disk stars are more enhanced in their $\alpha$ -element
abundances than the thin disk at a given [Fe/H] below solar metallicities. 
  
On the other hand, \citet{Ch00} 
have studied the chemical composition of 90 F and G disk dwarfs 
and did not find a clear separation
of [$\alpha$/Fe]vs[Fe/H] into thin and thick disks. 
As shown by \citet{Pr00},
this situation appears because the old high - metallicity thick 
disk stars with $T_\mathrm{eff} < 5700\ \mathrm{K}$
were not been included in the sample of dwarfs investigated by \citet{Ch00}.  

Recent investigations showed that 
abundance trends for Mn and Eu are also different for these two disk 
components.
In particular, a detailed study of the Mn abundance trend of the disk and 
metal-rich halo stars was carried out by \citet{Ni00}.
\citet{PW00}
have revisited these data (with the new hyperfine structure of 
the MnI line) and have shown that thin disk stars with 
-0.8 < [Fe/H] < -0.2  have [Mn/H]$\simeq$ -0.1
while thick disk stars [Fe/H] < -0.6 have [Mn/Fe]$\simeq$ -0.3.   
\citet{MG00,MG01} report that a step-like change exists in [Eu/Ba] and [Ba/Fe] 
ratios at the the transition of the thick to thin disk. 
They also found 
that europium is overabundant relative to barium and iron in the halo 
and thick disk stars. Nearly solar [Eu/Fe], [Ba/Fe] and [Eu/Ba] ratios 
are found for thin disk stars.  

In this paper, we consider the thick and thin disks as two separate components
of the Galactic disk which have been  formed at different epochs and have different 
evolutionary timescales.

\citet{MG86} suggested 
that the evolution of the abundance of oxygen vs iron can be explained by 
an origin of iron and oxygen from different SN types. 
Such an interpretation is widely accepted since provides an explanation 
for the observed overabundances of $\alpha$-elements
in the halo.
Iron is predominantly synthesized by type Ia SN
(intermediate mass stars in binary systems), while $\alpha$ elements
(in our case O, Mg, Si) are synthesized by massive stars.
Thus, abundance ratios depend on the stellar nucleosynthesis, stellar lifetimes
and initial mass function. 
But the variation of abundance ratios as a function of metallicity 
or time depends on the star formation history as well.
This allows us to use  the variations of [$\alpha$/Fe] 
as an indicator of the star formation history \citep{Ma92}.
Since the [$\alpha$/Fe] ratios  for the thick and thin disk stars are
distinct, we make the assumption  that star formation histories  
of the two disk components must be different.
The aim of this paper is to find  such parameters 
for the star formation history  
which allow us to reproduce the observed $\alpha$ - element abundances 
in the thin and thick disks of the Galaxy 
at metallicities below the solar value.

This paper is organized as follows.
In Sect.2 we describe the model  and its main components for the 
chemical evolution of the Galactic thin and thick disks.  
In Sect. 3 we show  and discuss the model predictions for the 
abundances of elements considered (Mg, Si, Fe)
for the thick and thin Galactic disks.  
In Sect. 4 we summarize the main conclusions. 
 
\section{The model}
Our model for the chemical evolution of the Galaxy 
is described in detail by \citet{Pil96} and \citet{Pil93}.
%The chemical evolution of Galactic disk is considered. 
We construct an open two-zone chemical evolution model
of the Galactic disk in order to investigate  the 
distinction in chemical characteristics of the Galactic 
disk subsystems. 
The Galactic disk is divided into two zones - 
the thin and thick disks; the mass of gas, heavy elements, stars  
and stellar remnants are computed as a function of time for each component.
The gas infall from the extrahalo takes place 
during their evolution.

In our model the continuous star formation process in the galactic disk 
is considered as a sequence of bursts with a generation of stars formed during 
each burst.
\footnote{This type of description of star formation rate is usually used 
for a numerical model. In our case "star fomation burst" means a number
of stars 
born in a short time interval. The advantage of using this 
type of description is that we can obtain any type of star formation history 
by varying the time interval between bursts and the burst amplitude. 
Such a description of the star formation history is rather general and 
allows us to produce a star formation history in the form of bursts as a 
continuous star formation history.} 
{Each stellar generation contributes to the chemical enrichment of the 
interstellar 
medium; the next stellar generation is formed from the galactic gas enriched 
by heavy elements ejected from the previous stellar generation.
We set an interval between bursts of 50 Myr as such a value is 
most suitable for a Galactic disk model. The number of bursts depends on 
the interval between bursts and  the duration of disk evolution.}
Thus, during galactic evolution the heavy element fraction 
and number of stars increases while the galactic gas mass decreases.
We have used use the equations (3)-(5) from 
\citet{Pil93} for a description of the temporal variations of the mass of gas, 
stars and heavy elements during galactic evolution. 
It is assumed that the evolution time of the disk of our Galaxy is 
13 Gyrs \citep{C91} and the burst interval has been set to  50 Gyr \citep{Pil96} .

The star formation rate $\psi(t)$ in thick and thin disks  
is described in the following way \citep{Pil96}:
\[ \psi(t)\sim\left\{      %%ia(r) e?a(r)¢áª ï sfr!!!%%%%%%%
\begin{array}{rl}
t\cdot e^{-t/T_{top}},& \mbox t \leq{}T_{top}\\
   e^{-t/T_{sfr}}, & \mbox t \geq{}T_{top}\nonumber 
\end{array}
\right. \] 
where $T_{top}$ and  $T_{sfr}$ are free parameters of the star formation rate.

The star formation history is described by a set of star formation bursts 
whose amplitudes are derived as
\begin{equation}
M_{b_{i}}=
\int_{t_{b_{i-1}}}^{t_{b_{i}}} 
\psi(t)\,dt,
\nonumber
\end{equation}
where  $t_{b_{i}}$ and $t_{b_{i-1}}$ are the times of beginning of 
j- and j-1  star formation bursts. The interval between consistent 
star formation bursts is 50 Myr.

We also assume that infall of intergalactic gas takes place on the disk  
during the galaxy's lifetime. According to \citet{Pil96}, the infall rate the
is described by the function  
$$A(t)=a_0{}e^{-t/T_{inf}}, $$ where $T_{inf}$ and $a_0$ are free model 
parameters.
The infalling gas has the primordial chemical composition
since, according to \citet{To88}, the accretion of a gas with metallicity 
up to 0.1 solar abundance gives the same result as accretion 
of a gas with primordial chemical composition. 

The star formation histories of the disk components were chosen 
so that the thick disk stars were formed
10-13 Gyr ago while the majority of the thin disk stars would have ages 
less than 10 Gyr.
\begin{table}
\caption{Parameters of selected models} 
\label{tab:1}
\begin{tabular}{ccc} \hline \hline
Model  & Thin disk & Thick disk \\
\hline
T$_{top}$[Gyr] & 1    & 1  \\
T$_{sfr}$[Gyr] & 8    & 5  \\
T$_{inf}$[Gyr] & 5    & 7  \\
a$_0$          & 0.06 & 0.1  \\
\hline
\end{tabular}
\end{table}

\begin{figure}
% Fig.1
%\resizebox{\hsize}{!}{\includegraphics{001fig.eps}}
\includegraphics[width=8.5cm,height=12cm]{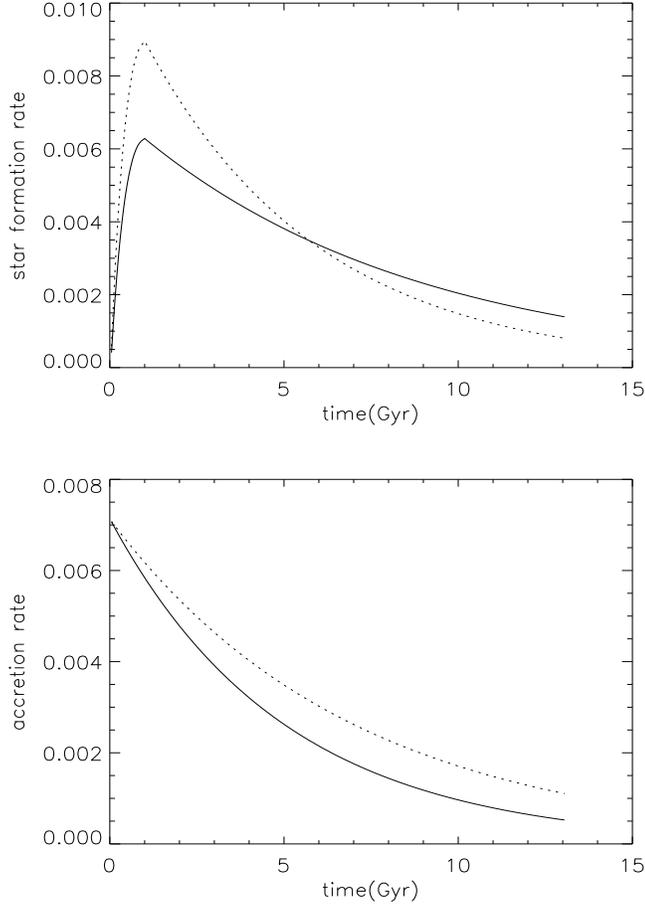}
\caption{The star formation and the accretion rates as a function of time  
in the model for thin (solid line) and thick (dashed line) disks.}  
\label{f:1}
\end{figure}

\subsection{Nucleosynthesis}

The abundance  of the  elements synthesized by a stellar generation and
ejected to the interstellar medium depends on 1) {}
the synthesis of elements by stars of various masses 2) {} the number
of stars formed in a given interval of stellar masses i.e. on the
initial mass function.
We adopted the yields of the Padova group \citep{Ma00,PCB98})
for calculation of the heavy element yields synthesized and ejected 
by massive, intermediate and low-mass stars 
during and at the end of stellar lifetimes.
The elements that have been included in the calculation are 
He,C,O,Mg,Si,S,Fe,Ca. 
The initial mass function is described
by the Salpeter law  with A=2.35 \citep{S55}.
In this paper we have used  the yields with above-solar initial metallicity,
Z=0.015.
Drastic differences exist between predicted yields for stars with
initial metallicity Z = 0 and Z > 0 \citep{WW95} therefore
the effect of a reasonable initial metallicity 
on final stellar yields is negligible for stars with Z > 0.

The Mg yield in the paper of \citet{PCB98} 
was determined from Woosley and Weaver's  yields \citep{WW95}
which give an underestimated Mg value  (see \citet{TGB00});
therefore the [Mg/Fe] ratio is not reproduced well in their paper
\citep{PCB98}.

\citet{Fra04} tested the Woosley and Weaver yields 
and found that the Mg yield is underestimated.   
In order to reproduce the Mg abundance there is a need 
to increase the predicted Mg yield  from stars with masses 
11 - 20 $M_{\odot}$ by a factor of 7 
while the yields from stars with masses of more then 20 $M_{\odot}$
need be lowered  about two-fold. 

Making use of Portinari's data in the calculation of the model of chemical 
evolution
we found that stars formed from a gas with the initial metallicity 
$Z<0.15$ give a lower Mg yield so as to reproduce the   
observed data. But stars with Z=0.15
give even higher Mg yields; we had to lower
the predicted Mg yield from the stars with 9-15 $M_{\odot}$ by a factor 
of 1.5 so that the model results would be in good agreement with the observation data
in the [Mg/Fe]vs[Fe/H] and [Si/Fe] vs[Mg/Fe] diagrams.

\begin{figure}
% Fig.0
\includegraphics[width=8.5cm,height=5.5cm]{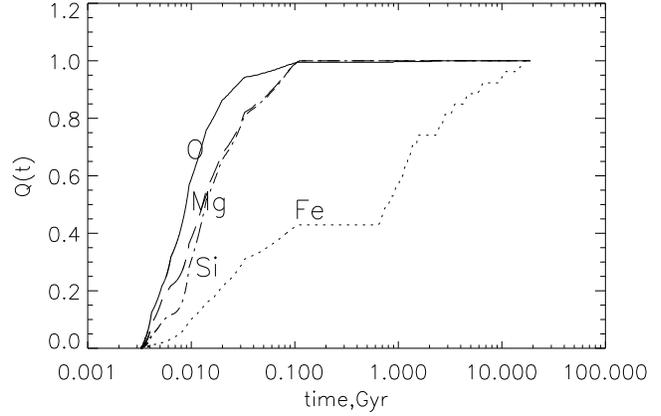}
%\resizebox{\hsize}{!}{\includegraphics{002fig.eps}}
\caption{The cumulative amounts of newly  synthesized elements
ejected by a stellar generation as a function of time.  
Each curve is normalized to its value at $t=13\times 10^{9}$ yrs.}  
\label{f:0}
\end{figure}
Fig.\ref{f:0} demonstrates the cumulative mass of magnesium, silicon, 
oxygen and iron synthesized by a single stellar generation. 
The iron yield contributed by SNIa is taken into account in the calculation 
of the total Fe yield. It was assumed that type Ia supernovae evolve 
according to a SD (single degenerate) scenario, in the framework of which 
the accretion of He or H on a white dwarf in a binary system 
is considered \citep[see][]{N82,HKN96}.
The Fe contribution  from  SNII and SNIa types at all times of evolution 
is about 1/3 and 2/3 correspondingly.
 
\section{The model results}

\subsection{The $\alpha$-element abundances}

\begin{figure}
% Fig.3
\includegraphics[width=8.5cm,height=18cm]{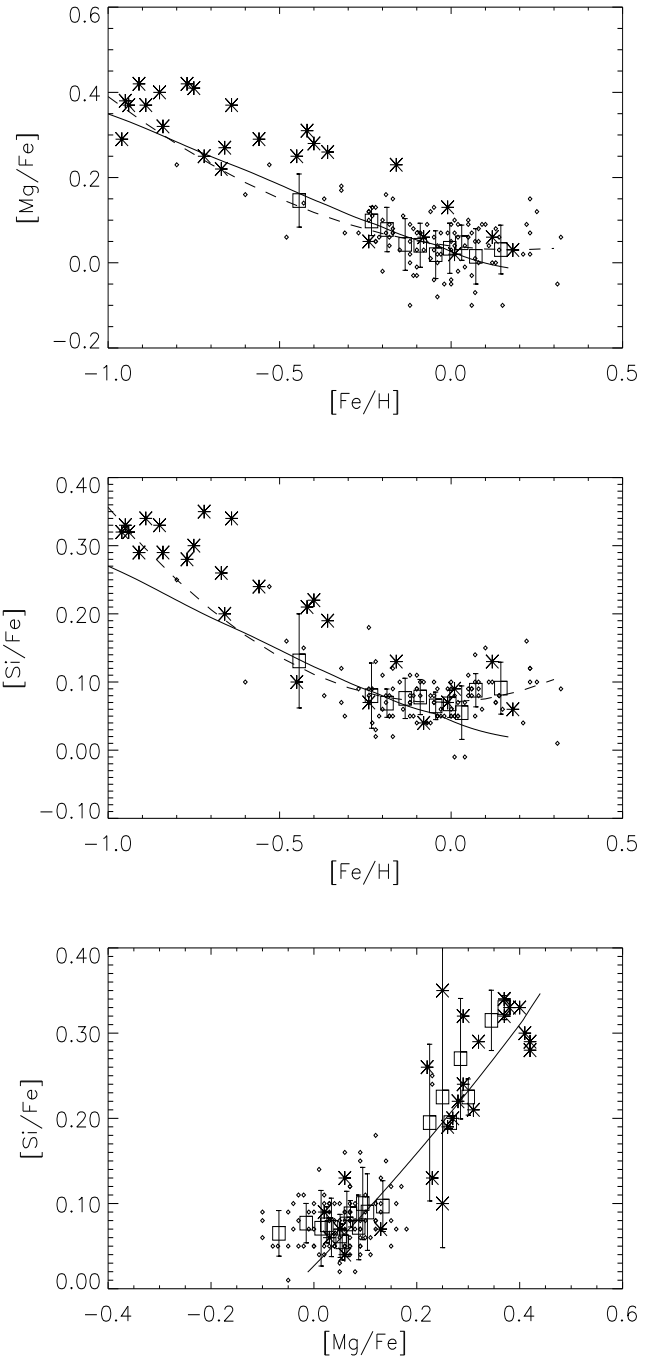}
%\resizebox{\hsize}{!}{\includegraphics{003fig.eps}}
\caption{The model predictions of the relative abundances of Galactic thin 
disk stars.
The data are taken from \citet{Mish04} and
\citet{BFL03}(thick disk star ages only).
The asterisks are thick disk stars,
the small diamonds are thin disk stars,
squares are observed data averaged in 10 bins with equal numbers 
of stars. The dashed line is a curve drawn by the least-squares method 
in the observed data. 
The model prediction for the thin disk is indicated by a solid line.} 
\label{f:3}
\end{figure}

\begin{figure}
% Fig.3a
\includegraphics[width=8.5cm,height=18cm]{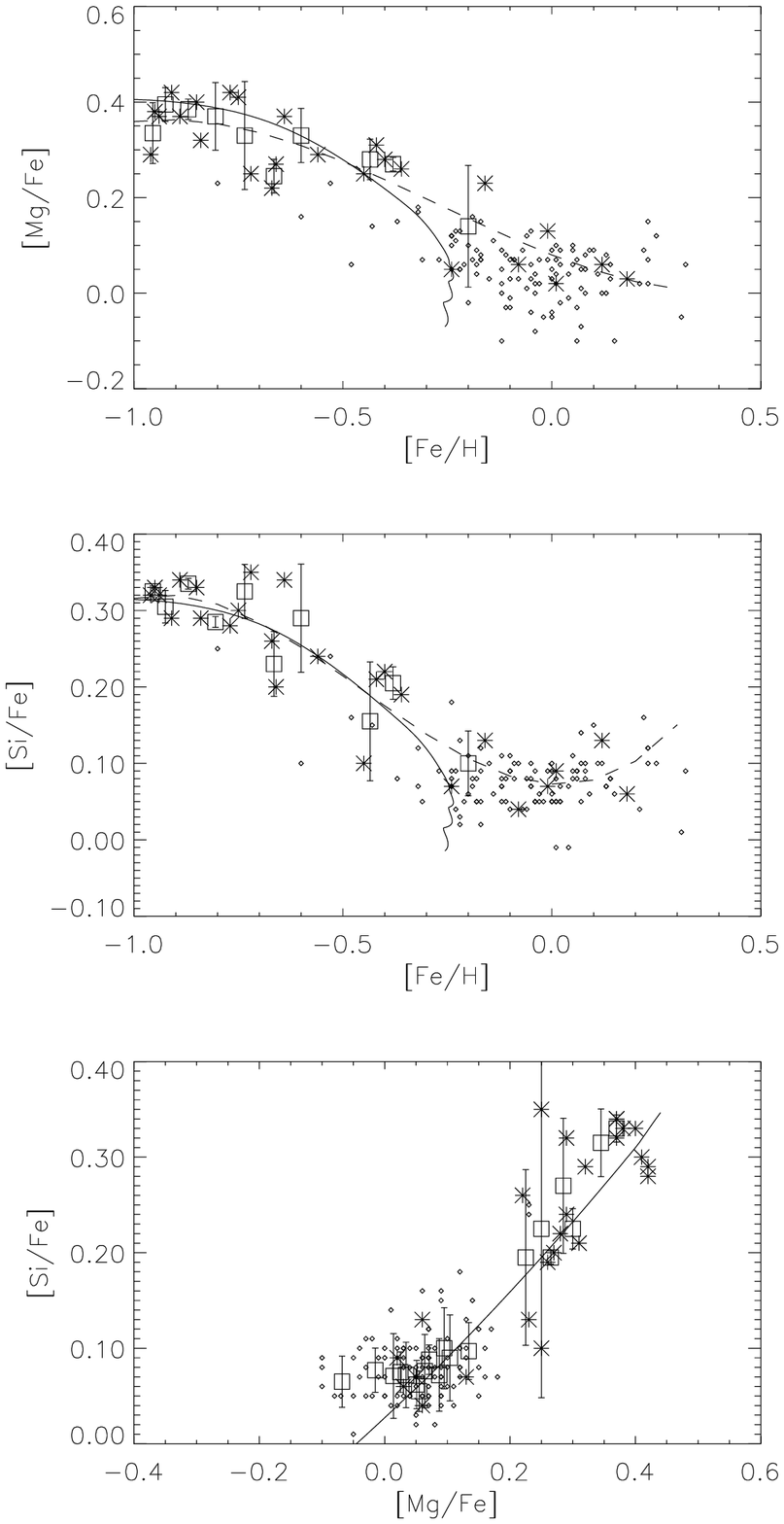}
%\resizebox{\hsize}{!}{\includegraphics{004fig.eps}}
\caption{The model predictions of the relative abundances of Galactic thick 
disk stars.
Symbols are the same as in Fig.\ref{f:3}.} 
\label{f:3a}
\end{figure}

In this paper we compared the model predictions with the observations  
of abundances of the thin and thick disk stars 
from the paper of \citet{Mish04}. Data of \citet{BFL03} (thick disk star ages)
were used in  addition for a more extended sample of thick disk stars.
In our paper, ages for Mishenina's data set were calculated  using  
the Bertelli isochrones \citep{Bert94}.

\begin{figure}
% Fig.31
\includegraphics[width=8.5cm,height=18cm]{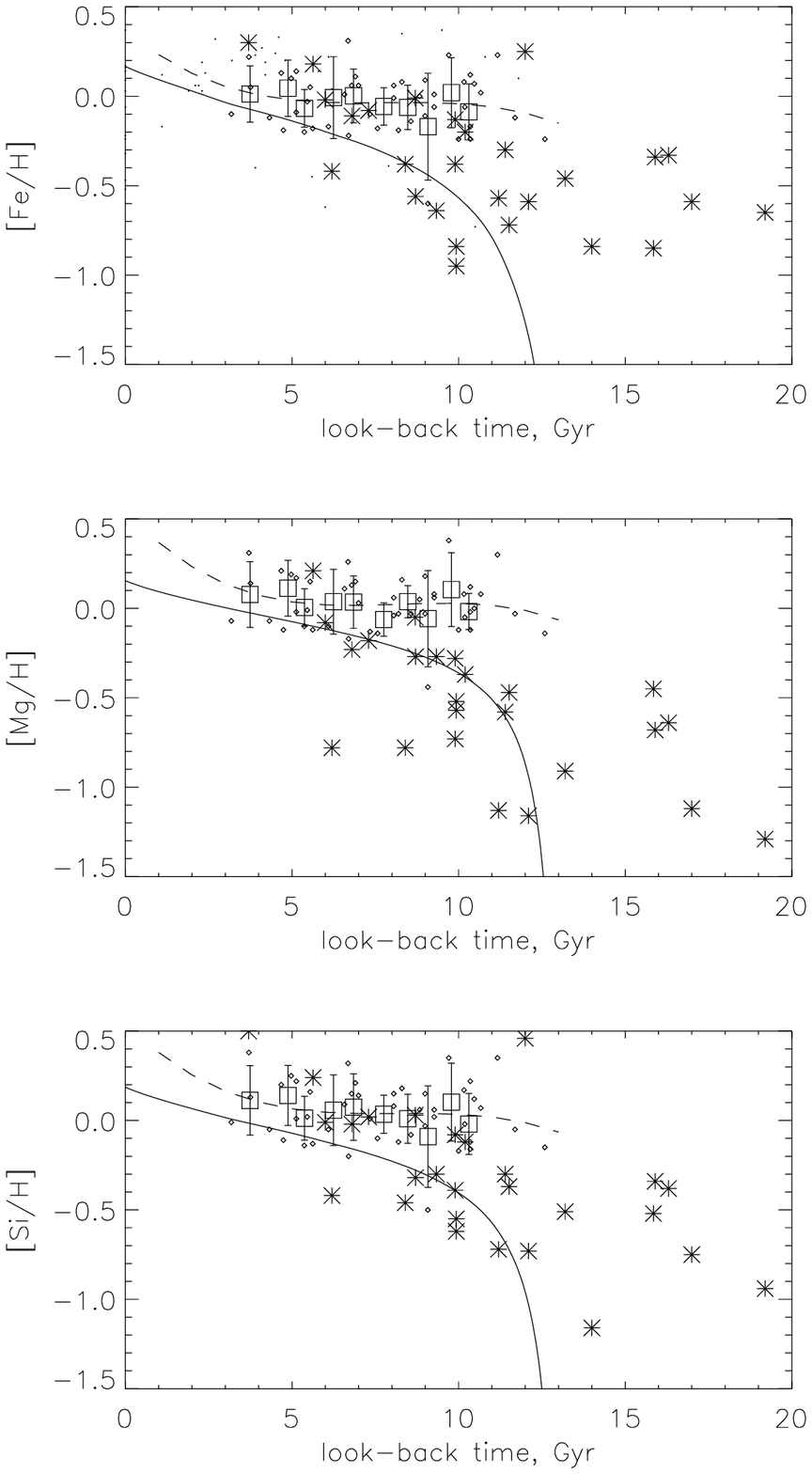}
%\resizebox{\hsize}{!}{\includegraphics{004fig.eps}}
\caption{The model predictions of the abundances of Galactic thin disk stars 
as a function of time.
Symbols are the same as in Fig.\ref{f:3}.} 
\label{f:31}
\end{figure}

The model predictions for the thin disk relative abundances are shown 
in Fig.\ref{f:3},\ref{f:31}. The best fit model parameters are 
listed in Table \ref{tab:1}.
The star formation and accretion rates for the thin disk are shown 
in Fig.\ref{f:1} a,b.

The parameters were chosen so that the
use of such parameters for the star formation and accretion rate 
would reproduce the observed data. 

\begin{figure}
\center
%\resizebox{\hsize}{!}{\includegraphics{006fig.eps}}
\includegraphics[width=8.5cm,height=5.5cm]{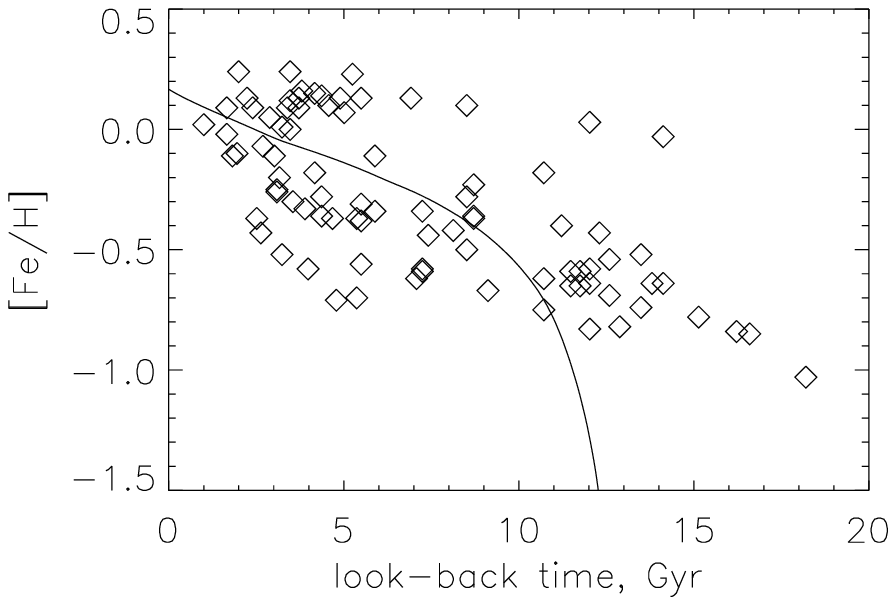}
\caption{The "age-metallicity" relation of the Galactic disk.  
Data are taken from \citet{edv93}. 
The model prediction of the thin disk is indicated by the solid line.} 
\label{f:4}
\end{figure}

As Fig.\ref{f:3} demonstrates, the use of
the above-mentioned parameter values allows us to 
reproduce the [Mg/Fe], [Si/Fe] and [Mg/Si] thin disk ratios quite well. 
The model track agrees closely with the line drawn by the least-squares  
method through a cloud of points marking the positions of the thin disk stars 
in the [Mg/Fe]vs[Fe/H] and [Si/Fe]vs[Fe/H] diagrams.
However, the model poorly reproduces the observed data  
at the super-solar metallicities especially for Si.  

Model predictions for the thick disk are shown in Fig.\ref{f:3a}.   
The best fit parameters of thick disk evolution are presented 
in Table \ref{tab:1}.
The star formation and accretion rates for the thick disk are shown 
in Fig.\ref{f:1} a,b.
As Fig.\ref{f:3a} shows, the solid line of the model prediction for 
the thick disk 
is in agreement with dashed line obtained by averaging the least-squares 
method of the observed relative abundances of Mg, Si and Fe. 

The model predictions for thin disk element abundances as functions of time 
are worse than for relative abundances. Fig.\ref{f:31} shows that the 
predicted abundance trends 
for the thin disk are slightly lower than Mishenina's 
observations averaged by the least-square method.

The obtained values of parameters of thin disk evolution 
are in good agreement with parameters of the best fit model of
the Galactic disk of \citet{Pil96} who have investigated the
"age - [Fe/H]" and "age - [O/H]"  ratios and
the solar neighbourhood metallicity distribution function.
The assumed disk age is 13 Gyr;
the obtained age - metallicity relation  reproduces quite well the 
observed disk star abundances of \citet{edv93}.
We compared the model prediction for the thin disk with
the observed abundances of disk stars obtained by \citet{edv93} 
for the whole disk. 
Fig.\ref{f:4} demonstrates that under the same parameters 
the thin disk modelled curve is in agreement with 
the observed age - metallicity relation. 

\begin{figure}
\center
%\vspace{5.5cm}
% Fig.6
\includegraphics[width=8.5cm,height=12cm]{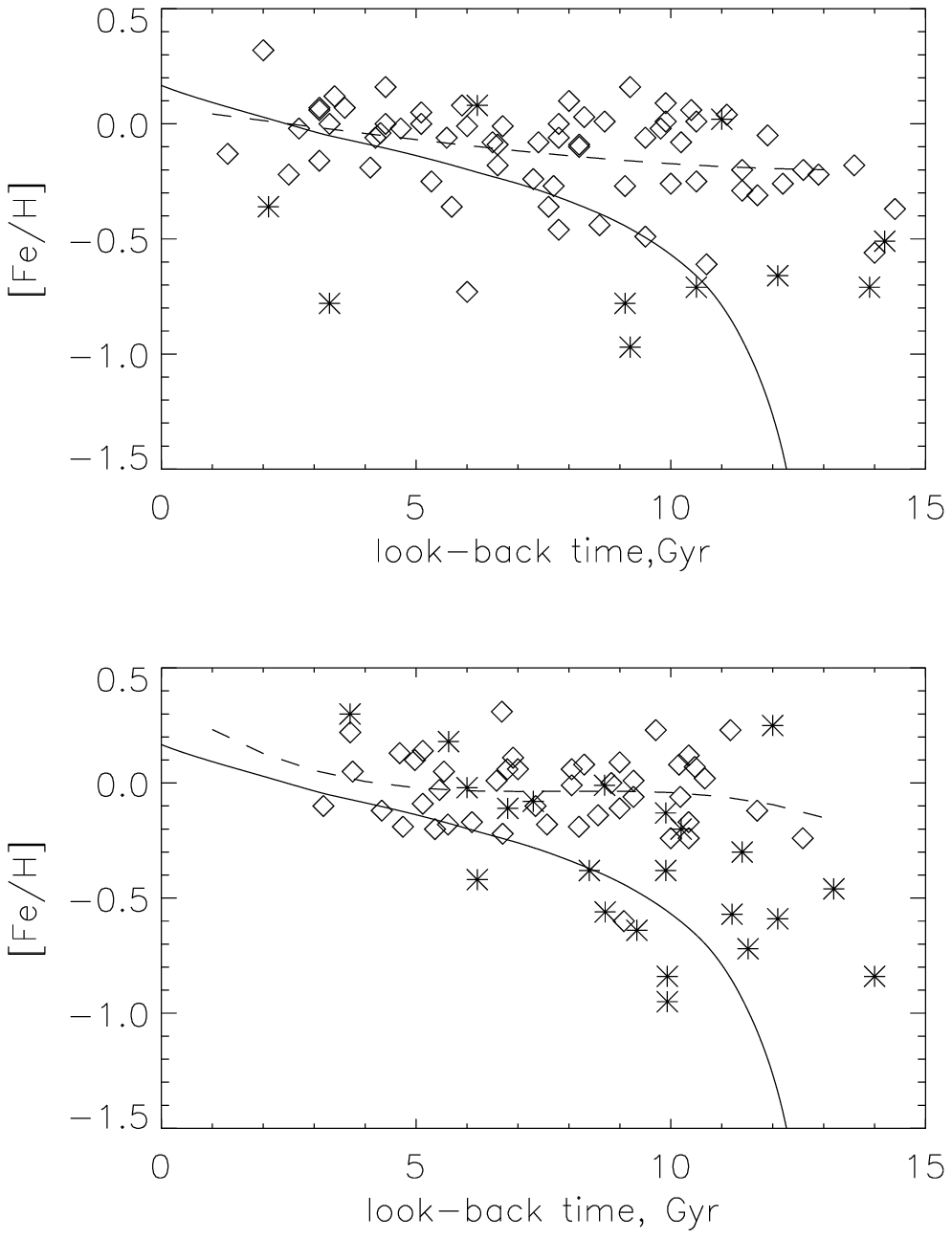}
%\resizebox{\hsize}{!}{\includegraphics{007fig.eps}}
\caption{ The "age - metallicity" relation of the thin and thick disk of the 
Galaxy obtained by \citet{Nord04}(top), 
and by \citet{Mish04}(bottom).
The thin disk stars are marked by large diamonds, 
thick disk stars - asterisks.  
The solid line is the model prediction for the thin disk, the dashed line 
is drawn by the least-squares method.} 
\label{f:6}
\end{figure}

Recently \citet{Nord04} presented  new determinations of metallicities and ages 
of F and G-dwarfs in the solar neighbourhood.
We used the  probabilities of the stars with measured velocities (U,V,W)
from the paper of \citet{Mish04}
and the ages and metallicities of Nordstr{\"o}m's survey 
for the stars investigated in both papers 
to build the new "age - metallicity" relation, Fig.\ref{f:6} (top).
Fig.\ref{f:6} shows that predicted abundances of the thin disk model 
($t_{gal}$ = 13 Gyrs) are also lower 
on average, as we can see in the case of Mishenina's data.

The computed probability of belonging to  
thin/thick disk for stars with measured velocities (U,V,W) of \citet{Mish04}
are in good agreement with the ones of \citet{BFL03}.  
\citet{Nord04} have compared their photometric metallicities
with the spectroscopic values for F and G-stars from 
\citet{edv93} and found that the agreement is excellent with a mean difference  
of only 0.02 dex and dispersion around the mean of 0.08.

Thus, there is the question of whether this "overestimated" 
(in comparison with the average metallicity of the whole disk)
mean value of abundances is usual for thin disk stars. 
If this is the case, it needs to provide a solution 
to this problem since it is not enough to change only star 
formation and accretion parameters in order to 
reproduce the thin disk age - metallicity relation
in the framework of our model.
{Such a age - metallicity relation (Fig. 5) can be explained
either by thin disk pre-enrichment or an infall of metal-enriched gas. 
We consider that in our model the thin disk is pre-enriched as 
we have used stellar yields with above-solar initial metallicity. 
\citet{To88} calculated the model with metal-rich gas infall 
for the Galactic disk. Table 2 in her paper demonstrates that the effect 
of a metal-rich infall  on the age - metallicity relation is negligible 
in the solar neighbourhood.
We ran a set of models with infall of enriched gas 
with Z from 0,1$Z_{\odot}$ to 1 $Z_{\odot}$. Solar proportions 
for the abundances of the various metals were assumed.  
We found out that the infall of enriched gas does not noticeably change 
the position of the modelled curve on the age - metallicity diagram.}

\begin{figure}
% Fig.3a1
\includegraphics[width=8.5cm,height=18cm]{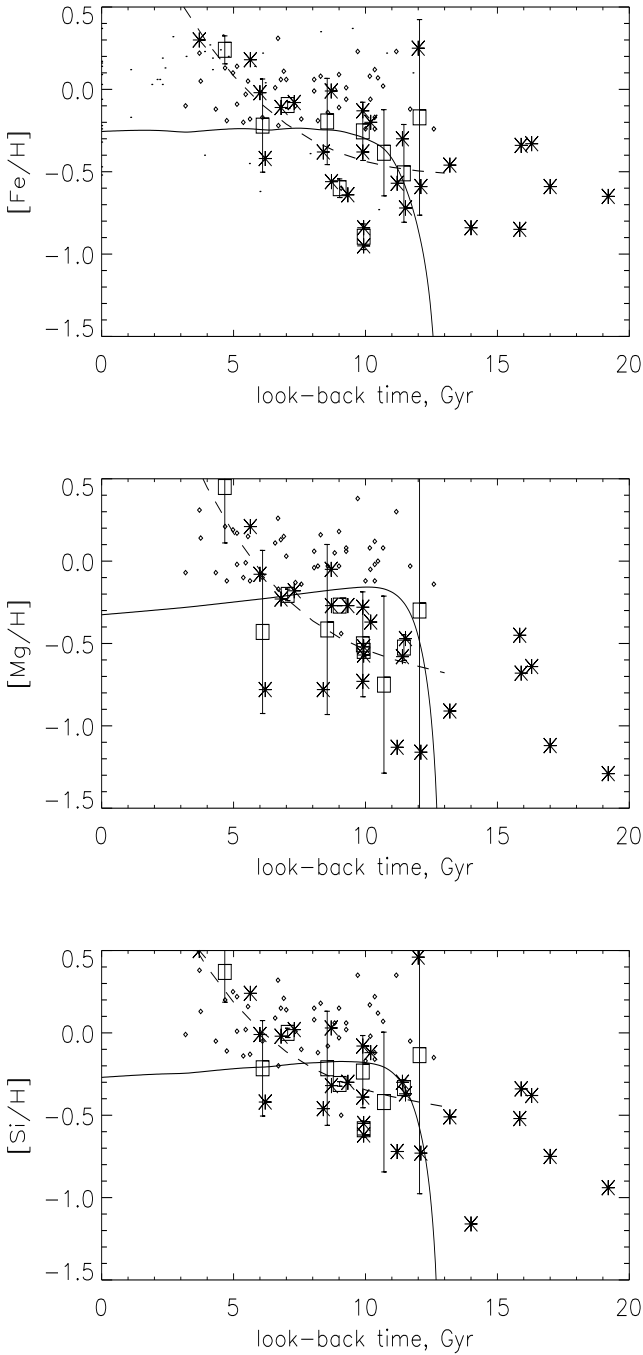}
%\resizebox{\hsize}{!}{\includegraphics{008fig.eps}}
\caption{The model predictions of the abundances of Galactic thick disk stars 
as a function of time.
Symbols are the same as in Fig.\ref{f:3}.} 
\label{f:3a1}
\end{figure}

Unfortunately, the number of stars belonging kinematically to the thin disk 
exceeds the number of stars belonging to the thick disk;
among the latter there are not enough objects for which one can reliably 
determine the ages. 
Therefore, the trend in the "age - metallicity" diagram for the thick disk 
stars is less reliable than for the thin disk stars. 
The few thick disk stars with determined ages 
were taken from \citet{BFL03,BFL04} 
to increase the number of thick disk stars 
in the "age- metallicity", "age- [Mg/H]" and " age - [Si/H]" diagrams.  

The situation is slightly complicated by the presence of stars with
thick disk kinematics and thin disk metallicity in the observed data
of \citet{Mish04}.
These stars are marked with asteriscs as belonging to the thick disk population
but are located in the region occupied by a younger
and more metal-rich thin disk population (e.g. Fig.\ref{f:3}).

The "age- metallicity" diagram (see Fig.\ref{f:3a}) shows that
the thick disk stars are less metal-rich and older than the thin disk stars.  
We have a problem determining the age range of thick disk stars. 
The use of various sets of isochrones 
gives ages of the oldest thick disk stars exceeding the age of Univers 
by several Gyrs.

Until the ages of the oldest stars are determined accurately,
we can consider the "age - metallicity" relation 
of thick disk stars as only 
an estimate of predicted abundance ratios. 
Nevertheless,  the predicted "age - metallicity" ratio  
(as "age -[Si/H]" ratio) reproduces the average values 
of the observed data for thick disk stars
marked by squares with errors in Fig.\ref{f:3a}. 

The abundance ratio of Mg and Si depends on the yields of single stars with 
different masses and lifetimes used in the model. 
As Fig.\ref{f:3} and \ref{f:3a} show,
the Mg and Si abundance ratios are in agreement with 
the observed data of the thin as thick disks,
except the observed data at super-solar metallicities.
The behavior of the thin and thick disk abundance trends at super 
solar metallicities requires further study. 

\subsection{The metallicity distribution function}

\begin{figure}
% Fig.2
\includegraphics[width=8.cm,height=5.5cm]{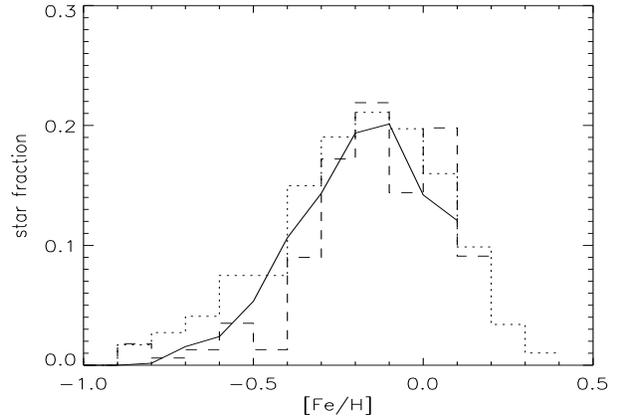}
%\resizebox{\hsize}{!}{\includegraphics{009fig.eps}}
\caption{The metallicity distribution function.
The dotted line is the observed distribution obtained by 
\citet{Hou98}, the dashed line is the observed distribution obtained 
by \citet{Jor00}, the solid line is the model prediction for the thin disk.} 
\label{f:7}
\end{figure}

The successful modeling of the chemical evolution 
should reproduce the G-dwarf metallicity distribution 
function of the Galactic disk.
The model prediction was compared 
with observations of the metallicity distribution function in
the solar neighbourhood (Fig.\ref{f:7}).
It is known that 94 $\%$ of solar neighbourhood stars
belong to the thin disk whereas the remaining 6 $\%$ belong to 
the thick disk population \citep{R96}.
Therefore the observed metallicity distribution functions in Fig. \ref{f:7}
were compared with the model prediction for the thin disk.
With the above-mentioned parameters of the star formation and gas 
accretion in the thin disk (see Table \ref{tab:1}),  
the model reproduces the observed metallicity distribution 
functions quite well (Fig.\ref{f:7}). 
 
\subsection{Comparison with previous work and discussion}

A model of chemical evolution should account for the origin and characteristics
of the subsystems of the Galaxy. A set of papers dedicated to
aspects of galactic chemical evolution exists in the literature.
Most of these papers deal with the standard open models of chemical evolution
with inflow of intergalactic gas in the Galaxy.

\citet{Ch97} considered a two-zone model 
for the disk and the halo, the two-infall model with two episodes of  
intergalactic gas infall,
in the formation process of the thin disk and the halo-thick disk.
Their model predicts the abundances of 16 chemical elements except
Mg and ${}^3$He. The authors conclude that the comparison between
theory and observations suggest a lower iron yield from SNIIs and an increased 
Mg yield from SNIIs. 
In the  framework  of Chiappini's model, \citet{Fra04} 
have computed  the chemical evolution of 12 elements 
(including O, Mg, Si) and have reproduced the 
results of observations for O, Si and Ca.
Their [Mg/Fe] ratio is underestimated owing to the low Mg yield from massive 
stars.  
\citet{Ch99} introduced the 
two-component model of chemical evolution 
which assumes that the thin and thick disk of the Galaxy have formed during 
two accretion episodes. Their models reproduce the metallicity distribution 
function, age - metallicity relation and [O/Fe]-[Fe/H] ratio
for the Solar vicinity.
\citet{ALC01} analyzed the evolution of all stable isotopes between
H and Zn. Their model also assumes the formation of the Galaxy in two
main episodes of exponentially decreasing infall. The first episode 
forms the halo and the  thick disk from infall of primordial extragalactic 
gas over 1 Gyr. The thin disk forms by infall of gas with metallicity 
0.12 Z$\odot$ over 7 Gyr. Their model predictions of abundance ratios 
are in agreement with observation except for Mg \citep[the authors used the 
yields of][]{WW95}. 

\citet{GP00} studied the evolution of the abundances of 
the intermediate mass elements ( C to Zn ) in the Galactic halo and local disk.
Their model assumes strong outflow in the halo phase and infall in the disk,
which allows them to reproduce the corresponding metallicity distribution.
The evolution of $\alpha$ - elements O, Si, S and Ca is well understood with
the assumption that SNIa contribute most of the Fe in the disk but
the Mg is underproduced in the \citet{WW95} yields, therefore their Mg yield
is relative to the observation in the disk. 
The model predictions cover a wide [Fe/H] range, 
from -4.0 up to 0.0, but do not show the clear limits of the model
components. One can define the limits only by comparing 
their predictions with observations. Unfortunately,
in the above-mentioned papers model predictions have not be compared 
with observations for the thin and thick disks separately
and we can only conclude about the evolution of the disk as whole.  

In contrast, we consider the thick disk evolution separately from the halo evolution. 
The new observations demonstrate 
the clear separation of Galaxy disk stars into the thin and thick disk 
population  in the range -0.6 < [Fe/H] < -0.3 (see Introduction) 
that shows the need to consider  the Galactic  
thick disk evolution separately, not only from the thin disk 
evolution but also from the halo evolution.    

\citet{Fer92} have considered the evolution of the solar neigbourhood 
as subdivided into halo and disk regions. The chemical abundances of  14 
species (in particular O, Mg, Si, Fe) have been predicted for the disk and 
halo. Their [O/Fe]vs[Fe/H] diagram demonstrates the predicted 
model curve covering the [Fe/H] range of -2.5 to 0.5 
(0.0 < [O/Fe] < 0.5)  for the disk
and -2.5 to -1. (0.2 < [O/Fe] < 0.5) for the halo. 
\citet{Tra99} have calculated the Galactic chemical evolution of 
the elements from Ba to Eu in the framework of the three-zone model of
chemical evolution (the halo, the thin disk, the thick disk).
Later, the Galactic evolution of Pb, Cu  and Zn was computed 
in framework of the same model \citep{Tra01,Mish02}.  
As an example of the model results, \citet{Tra99} 
computed the values of 
[O/Fe]vs[Fe/H] in the three Galactic zones of their standard model.
We can see from their Fig.3b that the thick disk phase covers 
the interval -2.5 $\lesssim$ [Fe/H] $\lesssim$ -1, and the thin disk phase 
begins at [Fe/H] $\gtrsim$ -1.5.
However, the new observations show clearly that the thick disk star population
continues up to [Fe/H]=-0.3 at least \citep{Mish04}.
According to \citet{Mish04}, star formation in the thick disk 
stopped when the enrichment was [Fe/H]=-0.3, [Mg/Fe]=+0.2 and [Si/Fe]=+0.17.
Our model predicts that the thick disk relative abundances (Fig.\ref{f:3a})
come abruptly to an end at [Fe/H]$\sim$-0.3 and deviate significantly 
from the least-square method curve at [Mg/Fe]$\sim$+0.2 and [Si/Fe]=+0.17.
In our case the thick disk star formation stopped at similar 
values as noticed by \citet{Mish04}.
But the observations show that there are a few thick disk stars 
with [Fe/H] > -0.3 \citep[see also][]{BFL05}. Our model does not predict 
the formation of these stars in the thick disk. It is possible that these stars 
were not formed in the thick disk  but were born in 
a galaxy - satellite and accreted into the Galactic disk.   

Looking at the observed ratios [O/Fe]vs[Fe/H] and
[Fe/O]vs[O/H] in the thick and thin disks, 
\citet{BFL04a} concluded that the thin and thick disks
were formed in different epochs and from a homogeneous gas.
In their view, the star formation rate  in the thick disk needs to be 
rapid to contribute 
the oxygen by SNII at the high metallicities before    
enrichment by SNIa (which do not contribute oxygen) will begin. 
Yet, the thin disk evolution must be slow and the star formation rate  
low. This conclusion is in good agreement with results of the two-infall 
model of \citet{Ch97}.Their model predicts a timescale of 1 Gyr for 
the thick disk and halo formation 
and about 8 Gyr for the thin disk formation.
Their model successfully  reproduced 
[$\alpha$/Fe] vs [Fe/H] for the Galactic thick disk-halo and the thin disk.
But they did not compare their results with observed thin and thick abundance 
disk trends  separately since at that time such observations had not been
obtained yet.
\begin{figure}
% Fig.2
%\resizebox{\hsize}{!}{\includegraphics[height=5cm]{010fig.eps}}
\includegraphics[width=8.cm,height=5.5cm]{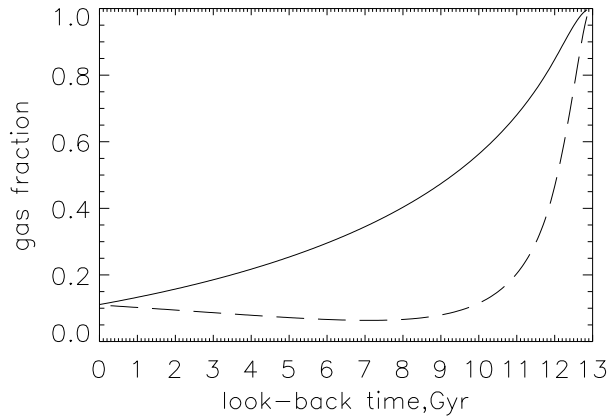}
\caption{The gas fraction as a function of time 
in the thin (solid line) and thick (dashed line) disk computed by the models.} 
\label{f:2}
\end{figure}

An indicator of star formation history in the model
is  the "gas fraction  - time" relation since it indicates 
what fraction of galactic gas was converted into stars and 
during what time it occured.
We found that the star formation history for the thick and thin disks 
must be different
to reproduce the observed abundance trends.
The obtained "gas fraction  - time" relation indicates that
the star formation history of the thick disk was rather brief as
the majority of the thick disk stars was formed 2 - 3 Gyrs
and had almost stopped more then 10 Gyrs ago (Fig.\ref{f:2}).
On the contrary, the thin disk stars only began to form 
9-10 Gyrs ago
(Fig.\ref{f:2}) and the star formation in the thin disk 
has decreased and is now almost stopped.

The accretion of intergalactic gas
significantly  affects the abundance predictions for the disk 
subsystems of our Galaxy. 
The observations show that the thin disk and thick disk exhibit 
parallel slopes of [$\alpha$/Fe] vs Fe/H] in the range 
-0.8 < [Fe/H] < -0.3. It is  supposed that the slope 
reflects the contribution of 
different supernovae to the interstellar medium enrichment.
In our paper, the Fe yield from SNIa is taken into account both for the thin and 
thick disks since the thick disk stars 
show evidence of chemical enrichment from SNIa \citep{Mish04,BFL03}.
The difference between the thick and thin disk abundance trends  
is caused by an increased inflow of gas mass in our model.
The trend of the relative abundances of the thin disk stars 
will change location  and approach the region occupied by the thick 
disk stars in the range of -1.0 <[Fe/H] < -0.3
if the mass of infalling gas at each accretion episode  
increases significantly (more than a factor of 10). 
In other words, the $\alpha$ - element enhancement of the thick disk as 
compared to the thin disk can be caused by more mass inflow in a unit 
of time in the thick disk.

\section{Conclusions}

The chemical evolution of the disk 
of the Galaxy was investigated in the framework of the two-zone open model with 
gas inflow.
It was supposed that the Galactic disk is divided into  two 
zones - thin and thick disks, differing from each other chemically 
and having different evolution timescales.

The  Galactic evolution  of
$\alpha$-elements (Mg and Si) and Fe was predicted  for the
thin and thick disks and was compared with recent observations. 
Our model predicts that the thick disk star formation stopped at [Fe/H]$\sim$-0.3 and at 
[Mg/Fe]$\sim$+0.2 and [Si/Fe]=+0.17; 
for the thin disk relative abundances
our model predictions cover the interval -1.0 < [Fe/H] < 0.0,
that agrees with observations. 

We have obtained the values of parameters of the star formation history 
for the thin and thick Galactic disks and conclude that 
it is necessary to use different star formation histories for 
the thick and thin disks to reproduce the observed abundance trends.    

Gas infall plays an important role in the appearance of $\alpha$ - element 
enhancement of the thick disk compared to the thin disk - 
the inflow rate must be more intensive per unit of time for the thick disk 
than the thin disk of the Galaxy.    

\acknowledgements{We thank   
Dr. L.S. Pilyugin, Dr. C. Travaglio  and  especially the referee, 
Dr. Fran\c cois, for helpful discussions and
useful comments.}

\bibliographystyle{aa}
\bibliography{myref}

\end{document}